\documentstyle[epsfig]{aipproc}

\def\beq{\begin{equation}}
\def\enq{\end{equation}}
\def\bea{\begin{eqnarray}}
\def\ena{\end{eqnarray}}
\def\Mesz{M\'esz\'aros}
\def\siml{\lower4pt \hbox{$\buildrel < \over \sim$}}
\def\simg{\lower4pt \hbox{$\buildrel > \over \sim$}}
\def\epm{\hbox{e}^\pm}
\def\L50{L_{w50}}
\def\E51{E_{w51}}
\def\cmcui{{\rm cm}^{-3}}
\def\tw{t_w}
\def\tw1{t_{w1}}

\def\rast{r_\ast}

\def\msun{M_\odot}

\newcommand{\boxsize}{0.89\textwidth}

\begin{document}
{\footnotesize
\noindent To appear in {\it Procs. 20th Texas Symp. on Relativistic Astrophysics, 2000}\hfill\\
{\it Eds. J. C. Wheeler, H. Martel (AIP Conf.Proc.), in press (2001)}
\bigskip

\title{GRB and Environment Interaction}

\author{ P. \Mesz$^{1}$ } 

\address{$^1$Pennsylvania State University, 525 Davey, University Park, PA 16802} 


\maketitle

\begin{abstract}

We discuss three aspects of the interaction between GRB and their surroundings. 
The illumination of the progenitor remnant and/or the surroundings by 
the X-ray afterglow continuum can produce substantial Fe K-alpha
line and edge emission, with implications for the progenitor model.
The presence of large dust column densities, capable of obscuring the GRB optical 
afterglow, will lead to characteristic delayed X-ray and far-IR light curve signatures.  
Pair production induced by the initial gamma-rays in the nearby environment will 
modify the initial spectrum and the afterglow light curve, and the magnitude of 
these changes provides a diagnostic for the external density. 

\end{abstract}

\maketitle

\section{Fe X-Ray Lines from GRB Progenitors}
\label{sec:feline}

Important clues for identifying the nature of the progenitors of the long
($t \simg 2$ s) GRBs may be available from the recent report at a 4.7$\sigma$
level of X-ray Fe line features in the afterglow after 1.5 days of the gamma-ray
burst GRB 991216  \cite{piro00}, as well as similar detections at the $3\sigma$
level in 5 other bursts with Beppo-SAX and ASCA. 
X-ray atomic edges and resonance absorption lines are theoretically expected to
be detectable from the gas in the immediate environment of the GRB, and in
particular from the remnants of a massive progenitor stellar system
\cite{mr98b,weth00,botfry00}.

A straightforward interpretation \cite{piro00} of the GRB 991216 observation
would imply a mass $\simg 0.1-1\msun$ of Fe at a distance of about 1-2 light-days,
possibly due to a remnant of an explosive event or supernova which occurred days
or weeks prior to the gamma-ray burst itself (a 'supranova', \cite{piro00,vi00}).
The long time delay is necessary both to get the relatively massive, slow moving
ejecta out to  few light-day distances (to explain the line appearance at
a few days with light travel arguments), and in order to get the initial Ni and
Co to decay to Fe ($\sim$ 55 days).  This requires a two-step process, in which
an initial supernova  leads to a temporarily stabilized neutron star remnant,
which after weeks collapses to a black hole leading to a canonical burst
(\cite{vs98,vi00}). It is unclear whether fall-back from the supernova leading
to the second collapse to a BH could occur with such a ($\sim$ weeks) long delay
(e.g. \cite{macfadyen00}). Another possibility is that a massive progenitor
has previously emitted a copious wind (${\dot M} \simg 10^{-4}\msun$/yr), which
would need to be unusally Fe-rich and highly inhomogeneous (\cite{weth00}; c.f.
\cite{piro00}).

An alternative, and perhaps less restrictive scenario for such Fe lines \cite{rm00}
involves an extended, possibly magnetically dominated wind from a GRB impacting the
expanding envelope of a massive progenitor star. This could be due either to a
spinning-down millisecond super-pulsar or to a highly-magnetised  torus around a
black hole (e.g. \cite{whee00}), which could produce a luminosity that was
still, one day after the original explosion,  as high as $L_m\sim 10^{47}
t_{day}^{-1.3}$ ergs. An outflow with such a dependence can also be powered by
accretion of fall-back material onto a central black hole \cite{macfadyen00}.
This jet luminosity may not dominate the continuum afterglow; but its impact on
the outer portions of the expanding stellar envelope at  distances
$\siml 10^{13}$ cm, even with just solar abundances, can be efficiently
reprocessed into an Fe line luminosity comparable to the observed value, 
together with a contribution to the X-ray continuum.  Under this interpretation, 
the dominant continuum flux in the afterglow, even in the X-ray band, is still 
attributable to a standard decelerating blast wave.

The relativistic magnetised wind from the compact remnant 
would develop a stand-off shock before encountering the envelope material, 
and shocked relativistic plasma would be deflected along the funnel walls.
Non-thermal electrons will be accelerated behind the standoff shock
in the jet material; the transverse magnetic field strength (which decreases as
$1/r$ in an outflowing wind) would be of order $10^4$ G at $10^{13}$ cm --
strong enough to ensure that the shock-accelerated electrons cool promptly,
yielding  a power- law continuum extending into the X-ray band. Some of these
X-rays would escape along the funnel, but at least half (the exact proportion
depending on the geometry and flow pattern) would irradiate the material in
the stellar envelope.
Pressure balance in the shoked envelope wall implies densities of $n_e =
\alpha L_m /6\pi r^2 c kT \sim 10^{17}\alpha L_{47} r_{13}^{-2}T_8^{-1}~\cmcui$,
where $\alpha\sim 1$ is a geometric factor, and the recombination time for
hydrogenic Fe in the funnel walls photoionized by the non-thermal continuum is
$t_{rec}=6\times 10^{-6}T_8^{1/2}n_{17}^{-1}
    \sim 6 \times 10^{-6} \alpha L_{m47}^{-1}r_{13}^2 T_8^{3/2}~\hbox{s}$.
Standard calculations of photoionization of optically-thin slabs (e.g. \cite{you99})
show that the equivalent width of the Fe K-alpha line, for solar abundances,
is about 0.5 kev, or twice as strong if the Fe has ten times solar abundances.
These results are applicable provided that the ionizing
photons encounter a Fe ion before being scattered by free electrons
i.e. provided that $\tau_T=\sigma_T d_i n_e \siml 1$.
Under these conditions  the Fe K-$\alpha$ photon flux is about 0.1 of
the X-ray continuum \cite{rm00},
$
{\dot N}_{LFe} \sim
 10^{54} L_{47} \beta ~{\rm ph/s},
$
where $\beta < 1$ is the ratio of ionizing to MHD luminosity.
This line luminosity compares well with Fe line luminosity $6\times 10^{52}$ ph/s
observed $t\sim 1.5$ day after the GRB 991216 burst by \cite{piro00}.

The total amount of Fe needed to explain the observed K-$\alpha$ line flux,
arising in a thin layer of the funnel walls of a collapsar model, amounts
to a very modest mass of $M_{Fe}\sim 10^{-8}\msun$, which could be Fe
synthesized in the core. The Fe-enriched core material can easily reach a
distance comparable to $r\sim 10^{13}$ cm  in 1 day for an expansion velocity
below the limit  $v\sim 10^9$ cm s$^{-1}$  inferred by \cite{piro00} from
the line widths. Even without this, a solar abundance ($10^{-5}\msun$ of Fe)
in the envelope is sufficient to explain the observations.
The initial, energetic portion of the relativistic jet, with a typical burst
duration of $1-10$ s, will rapidly expand beyond the stellar envelope, leading
in the usual way to shocks and a decelerating blast wave. A continually
decreasing fraction of energy, such as put out by a decaying magnetar, may
continue being emitted for periods of a day or longer,  and its reprocessing
by the stellar envelope can be responsible for the observed Fe line emission
in GRB 991216. Since the energy in this tail can decay faster than $t^{-1}$,
the usual standard shock gamma-ray and afterglow scenario need not be affected,
being determined by the first 1-10 s worth of the energy input.

\section{Dusty GRB Delayed XR/IR Afterglows}
\label{sec:xrdel}
 
For GRB in large star forming regions, a significant fraction of the prompt X-ray
emission will be scattered by dust grains. Since dust grains scatter X-rays by a
small angle, time delays of the scattered x-rays will be small (minutes to days,
depending on the X-ray energy and the grain size). If the dust column density is
substantial, the softer part of the X-ray afterglow on the above timescales will
be dominated by the dust scattering, the direct X-ray emission from the blast
wave being weaker. This intermediate time, soft(er) X-ray light curve will
be steeper than the unscattered X-ray afterglow.

As a specific example \cite{mg00}, consider a typical GRB whose unscattered
X-ray light curve is parametrized as
$F_0(t)=[1+(t/100{\rm s})]/[ 1+(t/100{\rm s})^{2.3}]$,
with an arbitrary normalization depending on the X-ray energy band.
This is represented by the thin line in Fig.\ref{fig:dust}.
We assume that the GRB occurs in a large star forming region, of typical
radius $R$ about 100pc, where the dust grain populations and optical
depths are close to what is observed in our Galactic center region.
Thus for numerical estimates we assume that (1) visual extinction is
$\sim 10$, (2) X-rays are scattered preferentially by those dust grains whose
size is in the range $a \sim 0.06\mu {\rm m}$, (3) the optical depth to
dust scattering at the X-ray energy $\epsilon$ is
$
\tau (\epsilon) = 3 \left( {\epsilon \over 1{\rm keV}}\right) ^{-2}.
$
At X-ray optical depths less than few, dust grains of size $a$ will scatter
X-rays of energy $\epsilon$ by an angle $\theta \sim 0.2\lambda /a$,
where $\lambda$ is the X-ray wavelength, $ \theta (\epsilon) \simeq
4\times 10^{-3}\left( {a \over 0.06\mu {\rm m}}\right) ^{-1}
\left( {\epsilon \over 1{\rm keV}}\right) ^{-1}$.
The time lag is $t\sim R\theta ^2/2c$, or
$
t(\epsilon )\sim 9\times 10^{4}{\rm s}\left( {a \over 0.06\mu {\rm m}}\right)^{-2}
 \left( {\epsilon \over 1{\rm keV}}\right) ^{-2}\left( {R \over 100{\rm pc}}\right).
$

At 2 keV, the optical depth is $\tau \sim 1$. The time lag is $t\sim 2\times 10^4$s.
The scattered flux is $F_s\sim \tau f/t\sim 0.03$. The unscattered flux at $2\times
10^4$s is $F_0\sim 10^{-3}$. In the time interval from hours to weeks, the dust
scattering dominates the afterglow, and, as shown in Fig.\ref{fig:dust}, the afterglow
is approximately a power law $F\propto t^{-1.75}$ \cite{mg00}. This is because dust
grains of
radius $a<0.06\mu {\rm m}$ will scatter the prompt emission with longer time lags,
$t\propto a^{-2}$, and with smaller optical depths $\tau$. To calculate $\tau$, we
take a standard dust grain size distribution where the number of grains of size of
order $a$ is $\propto a^{-2.5}$
For a scattering cross section $\propto a^4$
the optical depth is $\tau \propto
a^{1.5}\propto t^{-0.75}$, so the flux $F\propto t^{-1.75}$.
\begin{figure}[ht]
\begin{center}
\begin{minipage}[t]{0.5\textwidth}
\epsfxsize=\boxsize
\epsfbox{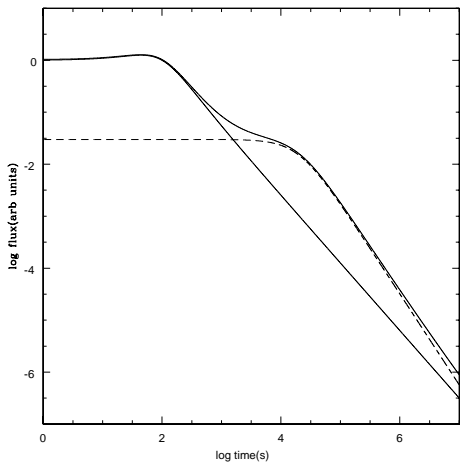}
\end{minipage}
\hspace{5mm}
\begin{minipage}[t]{0.4\textwidth}
\vspace*{-5cm}
\caption{\label{fig:dust} Dust-scattered X-ray afterglow. Thin line: unscattered
X-ray flux. Thin dashed line: scattered X-ray flux. Thick line: total flux. The flux
normalization is arbitrary, the relative fluxes correspond to the example
discussed in the text for an energy of 2 keV [6].}
\end{minipage}
\vspace*{-0.5cm}
\end{center}
\end{figure}

A GRB in such a highly obscured star-forming region should lead to specific
signatures in the X-ray afterglow, i.e. a bump in the X-ray light curve at 
energies $\epsilon \sim 2-3$ keV, hours to days after the burst \cite{mg00}.  
This X-ray signature is expected for bursts which do not produce
a detectable optical transient (OT).
Such OT-less, X-ray peculiar GRBs will also lead to thermal reemission
and scattering of the O/UV flux causing a delayed IR emission, as is the case
also for partially absorbed bursts \cite{wd00,eb00}.  For an isotropic
total burst energy $E\sim 10^{53}$ erg at a redshift $z\sim 1$ the
normalization of the X-ray flux for the burst of Figure 1 would be $F_x\sim 10^{-9}$
erg cm$^{-2}$ s$^{-1}$ keV$^{-1}$ for $t\siml 100$ s, in the usual range of X-ray
afterglow fluxes detected by Beppo-SAX. The dust reradiation
occurs beyond the sublimation radius $R_s\sim 10~L_{49}^{1/2}$ pc at wavelengths
$\lambda\simg 2(1+z)\mu$m, where $10^{49}L_{49}$ erg/s is the early UV component
of burst afterglow \cite{wd00}. The time delay associated with the
reradiated flux is $t_{IR} \sim (R_s/2c) \theta_j^2$ where $\theta_j=10^{-1}
\theta_{-1}$. 
At $z\sim 1$ the corresponding infrared flux at 2.2 $\mu$m would be
$F_{2.2\mu\hbox{m}} \sim L_{49} \theta_j^2 /[4\pi D_L^2 (R_s/2c)
\theta_j^2 \nu ]\sim 0.3 L_{49}^{1/2}$ $\mu$Jy, independent of $\theta_j$, or
$m_K\sim 23.3$ compared to Vega \cite{mg00}, approximately constant for a time
$t_{IR} \sim 5\times 10^6\theta_{-1}^{2}L_{49}^{1/2}$ s. 
Such $\gamma$-ray detected GRBs with anomalous X-ray afterglow behavior and no OT 
may be used as tracers of massive stellar collapses. It may thus be possible to 
detect star-forming regions out to redshifts larger than achievable with O/IR 
techniques, since typical GRB $\gamma$-ray, X-ray and IR fluxes can in principle 
be measured out to $z\sim 10-15$.

\section{Pair Production in GRB Environments}
\label{sec:pp}
 
Gamma-ray burst sources with a high luminosity can produce
$\epm$ pair cascades in their environment as a result of back-scattering
of a seed fraction of their original hard spectrum.  New pairs can be
made as some of the initial energetic photons are backscattered and
interact with other incoming photons. Previous work on this investigated
the acceleration of new pairs for a particular fireball model \cite{mt00,tm00},
the effect of pair formation for a low compactness parameter external shock
model of GRB \cite{db00}, and Compton echos produced by pairs \cite{mbr00}.
Here we  discuss a simplified analytical treatment \cite{mrr01} of pair effects
from $\gamma$-rays arising in internal shocks in a wind;
the remaining wind energy drives a blast wave which decelerates as it
sweeps up the external medium, and gives rise to the afterglow emission.
The  $\gamma$-rays would propagate ahead of the blast wave, leading to
pair production (and an associated deposition of momentum)  into the
external medium.  The pair cascades saturate after the external
(pair-enriched) medium reaches a critical bulk Lorentz factor, which is
generally below that of the original relativistic wind. For external
baryonic densities similar to those in molecular clouds the pairs can
achieve scattering optical depths $\tau_\pm \siml 1$.
Even for less extreme external densities the effect of the
additional pairs can be substantial, increasing the radiative efficiency
of the blast wave  and leading to distortions of the original spectrum.
This provides a potential tool  for diagnosing the compactness parameter of
the bursts and thus the radial distance at which shocks can occur.
It also provides a tool for diagnosing the baryonic density of the external
environment, and testing the association with star-forming regions.

For the maximum Lorentz factor to which an $\epm$ can be accelerated
by scattering, and the maximum  Lorentz factor at which back-scattered
photons can still make new pairs, one finds two regimes defined by
the effective duration of the light pulse seen by the screen of
accelerated pairs. At low radii (wind regime) the effective duration is 
the burst duration $t_w$; for large radii (impulsive regime),
the effective duration is $\Delta t \sim r/c\Gamma_\pm^2$.
For an incident photon number index
$\beta=2$, in the former $\Gamma_\pm\propto r^{-1/3}$ and in the latter
$\Gamma_\pm \propto r^{-2}$. The critical radius and Lorentz factor for
the transition between the wind and the impulsive regimes are \cite{mrr01}
$
r_{c}=  5\times 10^{14}\L50^{2/5}\tw1^{3/5}~~,~~
\Gamma_{c}= 3\times 10^1 \L50^{1/5}\tw1^{-1/5}.
$
The maximum radius at which pair cascades cut off is
$
r_{\ell}\sim (4 \rast c t_w /3 )^{1/2}\sim 4\times
10^{15}\L50^{1/2}\tw1^{1/2}~\hbox{cm}.
$
Before the pairs start accelerating, assuming they are held back by the
environmental protons through magnetic fields, an initial cascade amplification
fator $k_p\sim (m_p/m_e)$ is achieved. After the mean mass per scatterer
drops to a value comparable to the electron mass, before reaching $r_\ell$
a further amplification $k_a\sim 2^s \sim 50$ (where $s\sim log\Gamma_c/\log 2$)
is possible, so the total pair amplification factor is \cite{mrr01}
$k_c = k_p k_a(r_c) \sim (m_p/m_e) \Gamma_c \sim
  5\times 10^4 \L50^{1/5}\tw1^{-1/5}$.
The maximum pair optical depth at $r_c$, which is prevented from exceeding
$\tau_\pm \sim 1$ by self-shielding, is achieved for external densities
$n_p \simg n_{p,c}$, where
$
n_{p,c}\simeq 10^5 \L50^{-3/5} \tw1^{-2/5} \cmcui ~.
$

The external density and the initial Lorentz factor $\eta$
determine when the  outer shock and the reverse shock become important
and whether this happens within the radius already polluted with pairs.
If $\eta \siml r_l/c t_w$ the external shock
responsible for the afterglow occurs beyond the region ``polluted" by
new pairs, and otherwise the afterglow shock may experience,
after starting out in the canonical manner, a ``resurgence" or second
kick as its radiative efficiency is boosted by running into an
$\epm$-enriched gas \cite{mrr01}.  

Additional effects are expected when $\tau_\pm \to 1$, for
external baryon density $n_p \simg n_{c,p} \sim 10^5 \L50^{-2/5} \tw1^{-3/5} 
\cmcui$ at radii $r<r_\ell$. Such high densities could
be expected if the burst is associated with a massive star in which
prior mass loss  led to  a dense circumstellar envelope.  The pair optical
depth saturates to $\tau_\pm\sim 1$ and in addition to an increased
efficiency and softer spectrum of the afterglow reverse shock, the
original gamma-ray spectrum of the GRB will be modified as well.  
One of the consequences of such a critical external density leading to
$\tau_\pm\sim 1$ would be the presence of an X-ray quasi-thermal pulse,
whose total energy may be a few percent of the total burst energy \cite{mrr01}.

\bigskip\noindent
{This research was supported by NASA NAG5-9192, the Guggenheim
Foundation and the Sackler Foundation. I am grateful to M.J. Rees, A. Gruzinov
and E. Ramirez-Ruiz for valuable discussions on these topics.}

\end{document}